\documentclass[aps,prd,twocolumn,showpacs,showkeys,amsmath,amssymb]{revtex4}
\usepackage{epsfig} 
\usepackage{makecell}
\usepackage{array}
\usepackage{subfigure}
\usepackage{bm}
\usepackage{longtable}
\usepackage{tabularx}
\usepackage{booktabs}
\usepackage{appendix}
\usepackage{caption2}
\usepackage{amsfonts}
\usepackage{graphicx}
\graphicspath{{figs/}}
\usepackage{dcolumn}
\usepackage{siunitx}
\usepackage{bm}
\usepackage{times}
\usepackage[utf8]{inputenc}
\DeclareUnicodeCharacter{200B}{\space}
\usepackage{float}
\usepackage{txfonts}
\usepackage{overpic}
\usepackage{feynmf}   
\usepackage{epstopdf}   
\usepackage{slashed}  
\usepackage{multirow}

\usepackage[colorlinks, citecolor=blue,anchorcolor=red,menucolor=red, linkcolor=red,filecolor=red,runcolor=red,urlcolor=blue,frenchlinks=red]{hyperref}

\usepackage{ulem}
\usepackage{color,xcolor}

\allowdisplaybreaks[4]

\makeatletter
\newcommand{\figcaption}{\def\@captype{figure}\caption}
\newcommand{\tabcaption}{\def\@captype{table}\caption}

\newcommand{\Rmnum}[1]{\expandafter\@slowromancap\romannumeral #1@}

\def\hlinewd#1{%
  \noalign{\ifnum0=`}\fi\hrule \@height #1 \futurelet
   \reserved@a\@xhline}
\makeatother

\newcommand\dqq{\Big\langle \bar q q \Big\rangle}

\newcommand\dqGq{\Big\langle g_s \bar q \sigma G q \Big\rangle}

\newcommand\dGG{\Big\langle \alpha_{s} GG \Big\rangle}
\newcommand\dGGG{\Big\langle g_{s}^{3} G^{3} \Big\rangle}

\begin{document}
\title{Predictions of masses for light hybrid baryons}
\author{Qi-Nan Wang$^{1,\, 2}$}
\author{Ding-Kun Lian$^{1, \, 3}$}
\author{Wei Chen$^{1,\, 4}$}
  \email{chenwei29@mail.sysu.edu.cn (Corresponding author)}
\author{Hui-Min Yang$^{5}$} 
\author{Hua-Xing Chen$^3$}
\email{hxchen@seu.edu.cn}
\author{J. Ho$^6$}
\email{jason.ho@dordt.edu}
\author{T. G. Steele$^7$}
\email{tom.steele@usask.ca}
\affiliation{$^1$School of Physics, Sun Yat-Sen University, Guangzhou 510275, China \\
$^2$College of Physical Science and Technology, Bohai University, Jinzhou 121013, China\\
$^3$School of Physics, Southeast University, Nanjing 210094, China\\
$^4$Southern Center for Nuclear-Science Theory (SCNT), Institute of Modern Physics, 
Chinese Academy of Sciences, Huizhou 516000, Guangdong Province, China\\ 
$^5$School of Physics and Center of High Energy Physics, Peking University, Beijing 100871, China \\
$^6$Department of Physics, Dordt University, Sioux Center, Iowa, 51250, USA \\
$^7$Department of Physics
and Engineering Physics, University of Saskatchewan, Saskatoon, SK, S7N 5E2, Canada}
\thanks{Qi-Nan Wang and Ding-Kun Lian equally contribute to this work.}
\begin{abstract}
Within the method of parity-projected QCD sum rules, we study the mass spectra of light hybrid baryons with $I(J^{P})=1/2(1/2^{\pm}), 3/2(1/2^{\pm}), 1/2(3/2^{\pm}), 3/2(3/2^{\pm})$ by constructing the local $qqqg$ interpolating currents. We calculate the correlation functions up to dimension eight condensates at the leading order of $\alpha_{s}$. The stable QCD Lapalce sum rules can be established for the positive-parity $N_{1/2^+}, \Delta_{3/2^+}, \Delta_{1/2^+}$ and negative-parity $N_{1/2^-}, N_{3/2^-}, \Delta_{1/2^-}$ channels to extract their mass spectra. The lowest-lying hybrid baryons are predicted to be the positive-parity $N_{1/2^+}$ state around 2.01 GeV. These hybrid baryons mainly decay into conventional baryon plus meson final states. We propose to search for the light hybrid baryons through the $\Upsilon/\psi(3686)$ decays via the three-gluon emission mechanism in BESIII and BelleII experiments. Hopefully our studies of the light hybrid baryons will be useful for understanding the excited baryon spectrum and the behavior of gluonic degrees of freedom in QCD.  
\end{abstract}


\maketitle
{\it Introduction.}---As a distinctive prediction of QCD, the existence of hybrid states is a longstanding issue in both experimental and theoretical aspects in the past several decades. Different from the conventional $q\bar q$ meson and $qqq$ baryon~\cite{Gell-Mann:1964ewy,Zweig:1964jf}, hybrid states contain the gluonic degrees of freedom in the configurations, such as $\bar qgq$ hybrid mesons and $qqqg$ hybrid baryons. It is of crucial importance to identify the existence of hybrid states for understanding the nonperturbative properties of QCD~\cite{Meyer:2015eta,Chen:2016qju,Esposito:2016noz,Guo:2017jvc,Liu:2019zoy,Brambilla:2019esw,Chen:2022asf,Meng:2022ozq}.

In the past several decades, there are numerous investigations on hybrid mesons in various theoretical methods,  such as the MIT bag model~\cite{Barnes:1982tx,Chanowitz:1982qj}, Coulomb gauge QCD~\cite{Guo:2008yz}, LQCD~\cite{Lacock:1996ny,Lacock:1996vy,HadronSpectrum:2012gic,Ma:2020bex,Dudek:2011bn,Dudek:2013yja,Dudek:2009qf,Dudek:2010wm,Dudek:2011tt}, the flux tube model~\cite{Isgur:1984bm,Isgur:1985vy,Burns:2006wz}, Bethe-Salpeter equation~\cite{Burden:2002ps,Burden:1996nh}, QCD sum rules~\cite{Balitsky:1982ps,Latorre:1984kc,Govaerts:1983ka,Govaerts:1984bk,Govaerts:1984hc,Balitsky:1986hf,Chen:2013zia,Ho:2018cat,Ho:2019org,Palameta:2018yce,Barsbay:2022gtu,Chen:2021smz,Chen:2022qpd,Wang:2023whb,Lian:2024fsb,Wang:2024hvp} and so on. The most intriguing candidates for light hybrid mesons are $\pi_{1}(1400)$~\cite{IHEP-Brussels-LosAlamos-AnnecyLAPP:1988iqi}, $\pi_{1}(1600)$~\cite{E852:2001ikk},  $\pi_{1}(2015)$~\cite{E852:2004gpn} and $\eta_{1}(1855)$~\cite{BESIII:2022riz,BESIII:2022iwi} with exotic $J^{PC}=1^{-+}$. Comparing to hybrid mesons, a hybrid baryon is composed of a color octet triquark field and a color octet gluonic excitation. To date, the studies of hybrid baryons are much less appealing since they cannot carry exotic quantum numbers. In general, QCD allows arbitrarily strong mixing effect between $qqqg$ hybrid baryons and excited $qqq$ states with the same quantum numbers. 

In Refs.~\cite{Barnes:1982fj,Golowich:1982kx,Kisslinger:1995yw}, the Roper resonance $N(1440)$ was studied as a hybrid baryon with $J^{P}=1/2^{+}$. Such a possibility is excluded according to the meson-cloud theory from the constituent quark model~\cite{Burkert:2017djo}. In Ref.~\cite{Barnes:1982fj}, the masses of light hybrid baryon states with spin-parity $J^{P}=1/2^{+}, 3/2^{+}, 5/2^{+}$ were calculated in the MIT bag model. They predicted the masses of these hybrids in the range of $1.6-2.5 ~\mathrm{GeV}$, among which the lightest one is the $1/2^{+}$ state. In the flux-tube model, the lightest hybrid baryons are the mass-degenerate states with $J^{P}=1/2^{+}$ and $3/2^{+}$ near $1.87~\mathrm{GeV}$~\cite{Capstick:1999qq,Capstick:2002wm}. Using the relativistic quark model, the authors of Ref.~\cite{Gerasyuta:2002hg} studied the mass spectra of the nucleon and delta hybrids and suggested that $N(1710)$ and $\Delta(1600)$ could be hybrid baryons. In Refs.~\cite{Dudek:2012ag,Khan:2020ahz}, the mass spectra of positive-parity hybrid baryons were calculated in LQCD, in which $N_{1/2^+}$ and $N_{3/2^+}$ were predicted to be the lowest-lying hybrid states around $2.3-2.6~\mathrm{GeV}$. 
The QCD sum rule method has also been applied to study the hybrid baryons~\cite{Martynenko:1991pc,Kisslinger:1995yw,Kisslinger:2003hk,Azizi:2017xyx}. In Ref.~\cite{Martynenko:1991pc}, a light hybrid baryon $N_{1/2^+}$ was studied and found to be approximately $2.1~\mathrm{GeV}$. Using the same interpolating current, the calculation was improved to give different results of the gluon, trigluon and mixed condensates, while the hybrid mass was found to be around the observed mass of the Roper resonance~\cite{Kisslinger:1995yw}. In QCD sum rules, the $\Lambda(1600)$ and $\Lambda(1405)$ were suggested as strange hybrid baryons with $I(J^P)=0(1/2^{+})$~\cite{Kisslinger:2003hk} and $I(J^P)=0(1/2^{-})$~\cite{Azizi:2017xyx}, respectively. Recently, the triply heavy $cccg$ and $bbbg$ hybrids were studied in QCD sum rules~\cite{Zhao:2023imq} and a semirelativistic potential model~\cite{Cimino:2024bol}. 
 
In this work, we systematically study the mass spectra of nucleon and delta  hybrid baryons with both positive and negative parities by using the parity-projected QCD sum rules. 
We construct the $[qqq]_{8c}[G]_{8c}$ local hybrid baryon interpolating currents without the covariant derivative operators and introduce the parity projected sum rules to separate the hybrid baryons with different parities. We find stable mass sum rules and establish extensive mass spectra for the positive-parity $N_{1/2^+}, \Delta_{3/2^+}, \Delta_{1/2^+}$ and negative-parity $N_{1/2^-}, N_{3/2^-}, \Delta_{1/2^-}$ hybrid baryons. Our results show that the lowest-lying hybrid baryon is predicted to be the positive-parity $N_{1/2^+}$ state around 2.01 GeV. 
We propose to search for these light hybrid baryons through the $\chi_{cJ}/\Upsilon$ decays via the three-gluon emission mechanism in BESIII and BelleII experiments.

\begin{figure}[t!!]
  \centering
  \includegraphics[width=8cm,height=8cm]{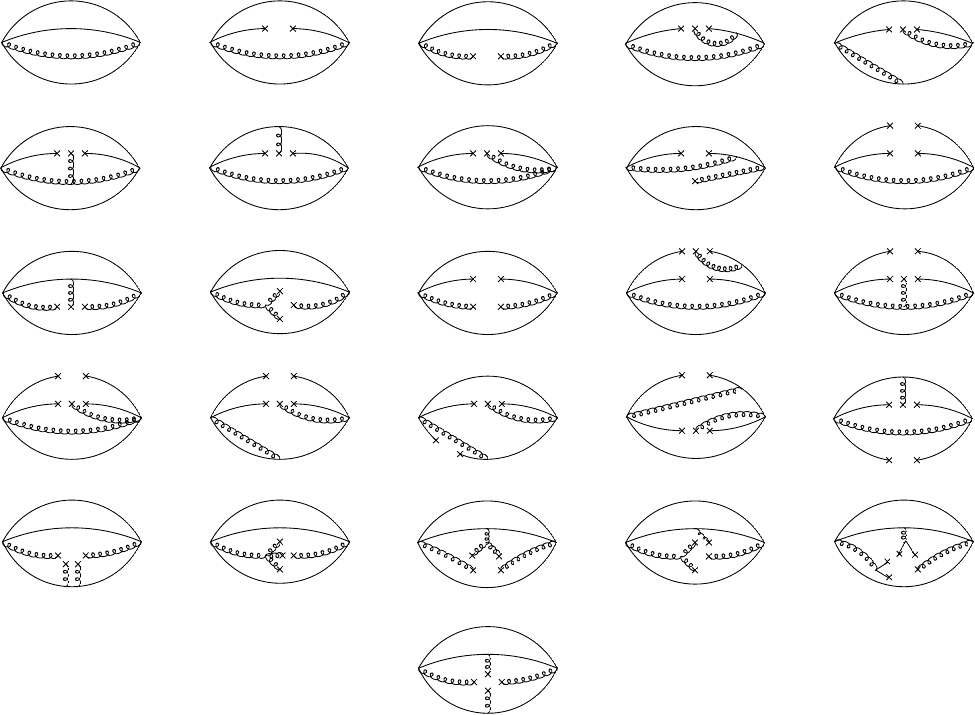}
\caption{Feynman diagrams involved in the calculations.}
  \label{fig: feyn_diagrams}
\end{figure}

{\it Hybrid baryon currents and parity-projected QCD sum rules.}---
A hybrid baryon current is composed of a color-octet triquark operator $[qqq]_{8c}$ and a color-octet excited gluon field. To investigate the low-lying states, we use the S-wave triquark fields introduced as ``Ioffe currents''~\cite{Ioffe:1981kw}. The gluon field is combined to compose the color-singlet hybrid baryon interpolating currents with $J^P=1/2^{+}$ and $3/2^{+}$ 
\begin{equation}
 \begin{aligned}  \label{Eq:currents}
    J_{1}&=\varepsilon^{abc}g_s[(u^{aT}C\gamma ^{\mu}u^{b})\gamma^{\nu}\gamma^{5}(G_{\mu \nu }d)^{c}-(u^{aT}C\gamma ^{\mu}d^{b})\gamma^{\nu}\gamma^{5}(G_{\mu \nu }u)^{c}]\, ,\\
    J_{2}&=\varepsilon^{abc}g_s[(u^{aT}C\gamma ^{\mu}u^{b})\gamma^{\nu}\gamma^{5}(G_{\mu \nu }d)^{c}+2(u^{aT}C\gamma ^{\mu}d^{b})\gamma^{\nu}\gamma^{5}(G_{\mu \nu }u)^{c}]\, ,\\
     J_{3}&=\varepsilon^{abc}g_s[(u^{aT}C\sigma^{\mu\nu}u^{b})\gamma^{5}(G_{\mu \nu }d)^{c}-(u^{aT}C\sigma^{\mu\nu}d^{b})\gamma^{5}(G_{\mu \nu }u)^{c}]\, ,\\
    J_{4}&=\varepsilon^{abc}g_s[(u^{aT}C\sigma^{\mu\nu}u^{b})\gamma^{5}(G_{\mu \nu }d)^{c}+2(u^{aT}C\sigma^{\mu\nu}d^{b})\gamma^{5}(G_{\mu \nu }u)^{c}]\, ,\\
     J_{5}&=\varepsilon^{abc}g_s[(u^{aT}C\gamma ^{\mu}u^{b})\gamma^{\nu}(\tilde{G}_{\mu \nu }d)^{c}-(u^{aT}C\gamma ^{\mu}d^{b})\gamma^{\nu}(\tilde{G}_{\mu \nu }u)^{c}]\, ,\\
    J_{6}&=\varepsilon^{abc}g_s[(u^{aT}C\gamma ^{\mu}u^{b})\gamma^{\nu}(\tilde{G}_{\mu \nu }d)^{c}+2(u^{aT}C\gamma ^{\mu}d^{b})\gamma^{\nu}(\tilde{G}_{\mu \nu }u)^{c}]\, ,\\
       J_{7}&=\varepsilon^{abc}g_s[(u^{aT}C\sigma^{\mu\nu}u^{b})(\tilde{G}_{\mu \nu }d)^{c}-(u^{aT}C\sigma^{\mu\nu}d^{b})(\tilde{G}_{\mu \nu }u)^{c}]\, ,\\
    J_{8}&=\varepsilon^{abc}g_s[(u^{aT}C\sigma^{\mu\nu}u^{b})(\tilde{G}_{\mu \nu }d)^{c}+2(u^{aT}C\sigma^{\mu\nu}d^{b})(\tilde{G}_{\mu \nu }u)^{c}]\, ,\\
    J_{1\mu}&=\varepsilon^{abc}g_s[(u^{aT}C\gamma ^{\nu}u^{b})(G_{\mu \nu }d)^{c}-(u^{aT}C\gamma ^{\nu}d^{b})(G_{\mu \nu }u)^{c}]\, ,\\
    J_{2\mu}&=\varepsilon^{abc}g_s[(u^{aT}C\gamma ^{\nu}u^{b})(G_{\mu \nu }d)^{c}+2(u^{aT}C\gamma ^{\nu}d^{b})(G_{\mu \nu }u)^{c}]\, ,\\
        J_{3\mu}&=\varepsilon^{abc}g_s[(u^{aT}C\gamma ^{\nu}u^{b})\gamma^{5}(\tilde{G}_{\mu \nu }d)^{c}-(u^{aT}C\gamma ^{\nu}d^{b})\gamma^{5}(\tilde{G}_{\mu \nu }u)^{c}]\, ,\\
    J_{4\mu}&=\varepsilon^{abc}g_s[(u^{aT}C\gamma ^{\nu}u^{b})\gamma^{5}(\tilde{G}_{\mu \nu }d)^{c}+2(u^{aT}C\gamma ^{\nu}d^{b})\gamma^{5}(\tilde{G}_{\mu \nu }u)^{c}]\, ,\\
      \end{aligned}
  \end{equation}
where $u/d$ is the up/down quark field, $G_{\mu \nu }$ is the gluon field strength, and $\tilde{G}_{\mu \nu}=\frac{1}{2}\varepsilon_{\mu\nu\alpha\beta}G^{\alpha\beta}$ the dual gluon field strength. The currents $J_{1,3,5,7,1\mu,3\mu }$ can couple to the nucleon hybrids $N_{1/2^\pm}, N_{3/2^\pm}$ with $I=1/2$, while $J_{2,4,6,8,2\mu,4\mu}$ couple to the delta hybrids $\Delta_{1/2^\pm}, \Delta_{3/2^\pm}$ with $I=3/2$. The couplings between currents and the corresponding positive-parity states can be written as~\cite{Chung:1981cc,Bagan:1993ii,Jido:1996ia,Ohtani:2012ps}
\begin{equation}\label{Eq:coupling_positive}
\begin{split}
  \langle 0|J_{i}|H_{+}(p)\rangle&=f_{+} u(p),\\
    \langle 0|J_{i\mu}|H_{+}(p)\rangle&=f_{+}^{\prime} u_{\mu}(p),
    \end{split}
  \end{equation}
in which $u(p)$ and $u_{\mu }(p)$ are the Dirac spinor and Rarita-Schwinger vector, respectively, and $f_+$ is the coupling constant. 
These currents can also couple to the negative-parity states via different coupling relations
\begin{equation}\label{Eq:coupling_negative}
\begin{aligned}
  \langle 0|J_{i}|H_{-}(p)\rangle&=f_{-}\gamma_{5} u(p),\\
    \langle 0|J_{i\mu}|H_{-}(p)\rangle&=f_{-}^{\prime} \gamma_{5} u_{\mu}(p).
    \end{aligned}
  \end{equation}
The two-point correlation functions can be written as
\begin{equation}\label{Eq:correlation}
  \begin{split}
\Pi(p^2)&=i\int  d^4x e^{ip\cdot x}\langle 0|T\left[J(x) \bar{J}(0)\right]|0\rangle=\Pi_{1/2}(p^2)\, ,\\
  \Pi_{\mu\nu }(p^2)&=i\int d^4x e^{ip\cdot x}\langle 0|T\left[J_{\mu}(x) \bar{J}_{\nu }(0)\right]|0\rangle\\
       &=-g_{\mu\nu } \Pi_{3/2}(p^2)+\cdots\, ,
  \end{split}
  \end{equation}
where $\Pi_{1/2}(p^2)$ and $\Pi_{3/2}(p^2)$ are the invariant functions for the hadron states with spin-1/2 and 3/2 respectively. They contain information of both positive- and negative-parity states
\begin{align} \label{Eq:correlationParity1}
  \Pi(p^2) &=\slashed{p}\Pi_{\slashed{p}}+\Pi_{I}\nonumber\\
  &=\frac{1}{2}\left(\frac{\slashed{p}}{\sqrt{p^{2}}}+1\right)\Pi_{+}(p^2)+\frac{1}{2}\left(\frac{\slashed{p}}{\sqrt{p^{2}}}-1\right)\Pi_{-}(p^2)\,
   \end{align}
in which $\slashed{p}=\gamma_\mu p^\mu$. $\Pi_{+}$ and $\Pi_{-}$ are the correlation function for positive- and negative-parity states respectively, which satisfy the following equation,
\begin{align} \label{Eq:correlationParity2}
  \Pi_{\pm}(p^2) &= \frac{f_{\pm}^{2}}{M_{\pm}^{2}-p^2}+\cdots,\,
\end{align}
where ``$\cdots$'' denote the excited states and continuum contributions. At the hadronic level, we use the dispersion relation to describe the invariant function as~\cite{Chen:2024ppj}
\begin{equation}
\Pi(p^{2})=\frac{(p^{2})^{N}}{\pi} \int_{0}^{\infty} \frac{\operatorname{Im} \Pi(s)}{s^{N}\left(s-p^{2}-i \epsilon\right)} d s+\sum_{n=0}^{N-1} b_{n}(p^{2})^{n}\, ,
\label{Cor-Spe}
\end{equation}
where $N$ is the number of subtractions required to make the integral convergent. The unknown subtraction terms shall disappear after applying the Borel transform. The spectral function can be defined in the ``narrow resonance'' approximation
\begin{align}\label{Eq:spectral}
  \rho(s) &\equiv \frac{\text{Im}\Pi (s)}{\pi   }
  =\slashed{p}\rho^{\slashed{p}}(s) + \rho^{I}(s)\\ \nonumber
  &=\frac{1}{2}(\frac{\slashed{p}}{\sqrt{s}}+1)f_{+}^{2}\delta (s-M_{+}^2)+\frac{1}{2}(\frac{\slashed{p}}{\sqrt{s}}-1)f_{-}^{2}\delta (s-M_{-}^2)+\cdots\,
\end{align}
where 
\begin{equation}
\begin{aligned}
\rho^{\slashed{p}}(s) & =\frac{1}{2}\frac{f_{+}^{2}}{\sqrt{s}}  \delta\left(s-M_{+}^{2}\right)+\frac{1}{2}\frac{f_{-}^{2}} {\sqrt{s}} \delta\left(s-M_{-}^{2}\right)+\cdots\, , \\
\rho^{I}(s) & =\frac{1}{2}f_{+}^{2}  \delta\left(s-M_{+}^{2}\right)-\frac{1}{2}f_{-}^{2}\delta\left(s-M_{-}^{2}\right)+\cdots\, .
\end{aligned}
\end{equation}
 The negative- and positive-parity states can be extracted as 
\begin{equation}
\begin{aligned}
\rho_\pm(s)=\sqrt{s}\rho^{\slashed{p}}(s)\pm\rho^{I}(s)\, .
\end{aligned}
\end{equation}
Then the QCD sum rules can be established as
\begin{equation}\label{Eq:PiFunction}
  \begin{split}
    \Pi_{\pm}(M_{B}^{2},s_{0})=\int _{(3m_q)^2}^{s_{0}} \rho_\pm(s) e^{-s/M_{B}^{2}}ds=f_{\pm}^{2}e^{-M_{\pm}^2/M_{B}^{2}} \, ,
     \end{split}
\end{equation}
where $\rho_{\pm}(s)=\text{Im}\Pi_{\pm}(s)/\pi$, $m_q$ is the light quark mass ($q=u/d$), $M_B$ and $s_0$ are the Borel mass and continuum threshold respectively. It is clear that $\rho_{\pm}(s)$ contains only contributions from positive/negative-parity baryon states. 
The lowest-lying hadron mass can be obtained as
\begin{equation}
   \begin{aligned}
  M_{\pm}\left(M_B^{2},s_{0}\right) & =\sqrt{\frac{\frac{\partial}{\partial\left(-1 / M_B^2\right)} \Pi_{\pm}\left(M_B^{2},s_{0}\right)}{\Pi_{\pm}\left(M_B^{2},s_{0}\right)}} \,. 
 \label{mass}
 \end{aligned}
\end{equation}
At the QCD side, the correlation functions and the spectral functions can be evaluated via the operator product expansion (OPE). In this work, we calculate the correlation functions at the leading order of $\alpha_{s}$ up to dimension-eight condensates, including the Feynman diagrams in Fig.~\ref{fig: feyn_diagrams}.
As an example, we show the spectral function for the current $J_7$ as
\begin{align} \label{Eq:Result}
\rho_{7}^{\slashed{p}}(s)&= \frac{3 \alpha_ss^4}{20480 \pi ^5}+\frac{\alpha_s\dqq^2 s}{12 \pi }-\frac{49 \dGG^2}{2304 \pi ^2}+\frac{11 \dGGG M^2}{3072 \pi ^4}\nonumber\\&-\frac{47\alpha_s\dqGq \dqq}{576 \pi }-\frac{93 \alpha_s^2 \dqq^2 M^2}{256 \pi ^2}\nonumber\\&+\frac{19 \alpha_s^2 \dqq^2 M^2\log \left(\frac{s}{\mu ^2}\right)}{64 \pi ^2}+\frac{13 \dGG^2 \log \left(\frac{s}{\mu ^2}\right)}{1152 \pi ^2}\, ,\\\nonumber
\rho_{7}^{I}(s)&= -\frac{\dGG \dqq s}{32 \pi }+\frac{55 \alpha_{s} \dqGq s^2}{768 \pi ^3}-\frac{5 \alpha_{s} \dqq s^3}{192 \pi ^3}\, .\\
\end{align}

At the leading order of $\alpha_{s}$, the dimension-4 and dimension-6 gluon condensates give no contribution for the spin-1/2 currents $J_{1}-J_{8}$. 
We calculate the $\dqq\dqq$ and $\dqq\dGG$ condensates by applying the factorization assumption, which is usually adopted in QCD sum rules to estimate the values of the high dimension condensates~\cite{Shifman:1978bx,Reinders:1984sr}.
There are eleven independent gluon condensates for dimension-8 operators\cite{Bagan:1984zt,Narison:2011xe}
\begin{align}
\phi_{1} &\equiv \langle \operatorname{Tr}(G_{\nu\mu}G^{\nu\mu}) \operatorname{Tr}(G_{\tau\rho}G^{\tau\rho}) \rangle, \nonumber \\
\phi_{2} &\equiv \langle \operatorname{Tr}(G_{\nu\mu}G^{\rho\mu}) \operatorname{Tr}(G^{\nu}{}_{\tau}G_{\rho}{}^{\tau}) \rangle ,\nonumber \\
\phi_{3} &\equiv \langle \operatorname{Tr}(G_{\nu\mu}G^{\tau\rho}) \operatorname{Tr}(G^{\nu\mu}G_{\tau\rho}) \rangle ,\nonumber \\
\phi_{4} &\equiv \langle \operatorname{Tr}(G_{\nu\mu}G^{\tau\rho}) \operatorname{Tr}(G^{\nu}{}_{\tau}G^{\mu}{}_{\rho}) \rangle ,\nonumber\\
\phi_{5} &\equiv  \langle \operatorname{Tr}\left(G_{\mu \nu} G_{\mu \nu} G_{\alpha \beta} G_{\alpha \beta}\right)\rangle\, ,\nonumber\\
\phi_{6} &\equiv \langle \operatorname{Tr}\left(G_{\mu \nu} G_{\alpha \beta} G_{\mu \nu} G_{\alpha \beta}\right)\rangle\, ,\nonumber\\
\phi_{7} &\equiv \langle \operatorname{Tr}\left(G_{\mu \nu} G_{v \alpha} G_{\alpha \beta} G_{\beta \mu}\right)\rangle\, ,\nonumber\\
\phi_{8} &\equiv \langle \operatorname{Tr}\left(G_{\mu \nu} G_{\alpha \beta} G_{\nu \alpha} G_{\beta \mu}\right)\rangle\,  ,\nonumber\\
\phi_{9} &\equiv\langle f_{a b c} G_{\mu \nu}^a J_\mu^b J_\nu^c\rangle\, ,\nonumber\\
\phi_{10} &\equiv\langle J_\mu^a D^2 J_\mu^a\rangle\, ,\nonumber \\
\phi_{11} &\equiv\langle f^{a b c} G_{\mu \nu}^a G_{\nu \lambda}^b D^2 G_{\lambda \mu}^c\rangle\, 
\end{align}
where  $J_\mu^a=g_{s}\sum_{u, d, s} \bar{q} \gamma_\mu\left(\lambda^a / 2\right) q$. We also apply the factorization assumption for these operators to simplify the calculations, as discussed in Refs.~\cite{Bagan:1984zt,Narison:2011xe,Huang:2014hya}
\begin{align}
g_{s}^{4}\phi_{1} &= \frac{1}{4} \langle \dGG^{2} \rangle^{2}\, ,\nonumber\\
g_{s}^{4}\phi_{3} &= -\frac{1}{16} \langle \dGG^{2} \rangle^{2} + 2g_{s}^{4} \phi_{2}\, ,\nonumber\\
g_{s}^{4}\phi_{4} &= -\frac{1}{32} \langle \dGG^{2} \rangle^{2} + g_{s}^{4}\phi_{2}\, ,\nonumber\\
g_{s}^{4}\phi_{5}&=\frac{ 16\pi^{2}}{12}\dGG^{2} \, ,\nonumber\\
g_{s}^{4}\phi_{6}&=-\frac{5\times16\pi^{2}}{48}\dGG^{2}+2g_{s}^{4}\phi_{2} \, ,\nonumber\\
g_{s}^{4}\phi_{7}&=g_{s}^{4}\phi_{8}=\frac{16\pi^{2}}{192}\dGG^{2}+\frac{1}{2}g_{s}^{4}\phi_{2} \, ,\nonumber\\
g_{s}^{3}\phi_{9}&=-\frac{3}{2}g_{s}^{4}\dqq^{2}M^{2} \,  ,\nonumber\\
g_{s}^{2}\phi_{10}&=-\frac{4}{3}g_{s}^{4}\dqq^{2}M^{2} \, ,\nonumber\\
g_{s}^{3}\phi_{11}&=\dGGG M^{2} \, 
\end{align}
where $g_{s}^{4}\phi_{2}$ is chosen as $\frac{7\times16\pi^2}{96}\dGG^{2}$\cite{Bagan:1984zt} and $M^{2}=0.3~\mathrm{GeV}^2$\cite{Nikolaev:1982ra}. 

In Eq.~(\ref{mass}), one should find reasonable working regions of $s_{0}$ and $M_B^{2}$ to determine the hadron mass. We require a good OPE convergence to obtain the lower bound of $M_{B}^{2}$
\begin{equation}
R_{D>6}=\left|\frac{\Pi_{\pm}^{D>6}\left(M_B^{2}, \infty\right)}{\Pi_{\pm}^{t o t}\left(M_B^{2}, \infty\right)}\right| \, ,\label{Eqs:Conver}
\end{equation} 
where $\Pi_{\pm}^{D>6}\left(M_B^{2}, \infty\right)$ represents the contributions from $D>6$ condensates. 
The upper bound of $M_B^{2}$ and the optimal value of $s_{0}$ are found by ensuring a sufficiently large pole contribution
 \begin{equation}
 P C=\frac{\Pi_{\pm}\left(M_B^{2}, s_0\right)}{\Pi_{\pm}\left(M_B^{2}, \infty\right)}  \, .\label{Eqs:PC}
\end{equation}

{\it Numerical analyses.}---To perform numerical analyses, we use the following values for various QCD parameters at the renormalization scale $\mu=1 ~\mathrm{GeV}$ and $\Lambda_{QCD}=300 ~\mathrm{MeV}$~\cite{ParticleDataGroup:2022pth,Narison:2011xe,Narison:2018dcr,Jamin:2002ev}:
\begin{align}\label{parameters}
m_u&=m_d=m_q=0\, ,\nonumber \\
\alpha_s \left(\mu\right) & =\frac{4 \pi}{29/3 \ln \left(\mu^{2} / \Lambda_{\mathrm{QCD}}^2\right)}\, , \nonumber \\
\dqq & =-(0.24 \pm 0.01)^3 ~\mathrm{GeV}^3 \,,  \nonumber\\
\dqGq & =(0.8 \pm 0.2) \times \dqq ~\mathrm{GeV}^2 \,,  \\
\dGG & =(6.35 \pm 0.35) \times 10^{-2} ~\mathrm{GeV}^4 \,,  \nonumber\\
\dGGG & =(8.2 \pm 1.0) \times \dGG ~\mathrm{GeV}^2  \, .\nonumber
\end{align}

\begin{figure}[h!!]
  \centering
  \subfigure[]{\includegraphics[width=0.23\textwidth]{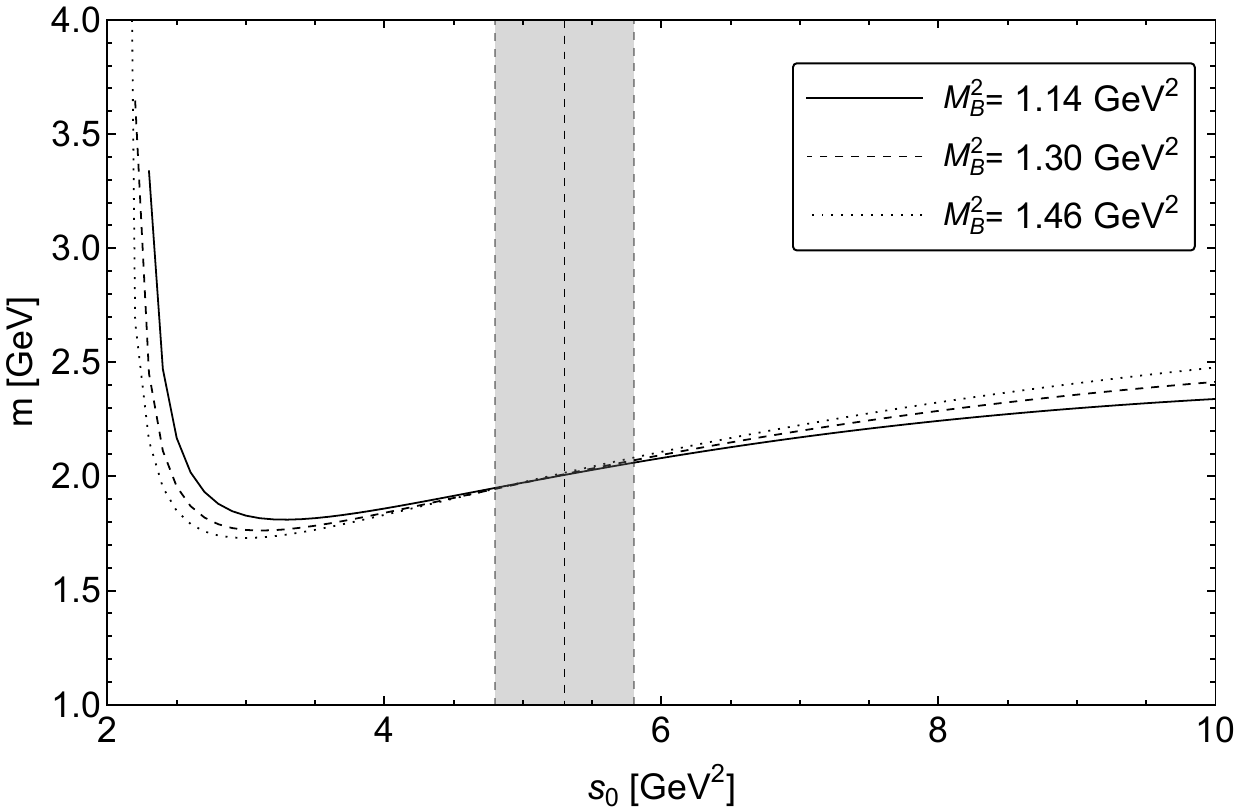}\label{fig:J7-s0-m-plus}}
  \subfigure[]{\includegraphics[width=0.23\textwidth]{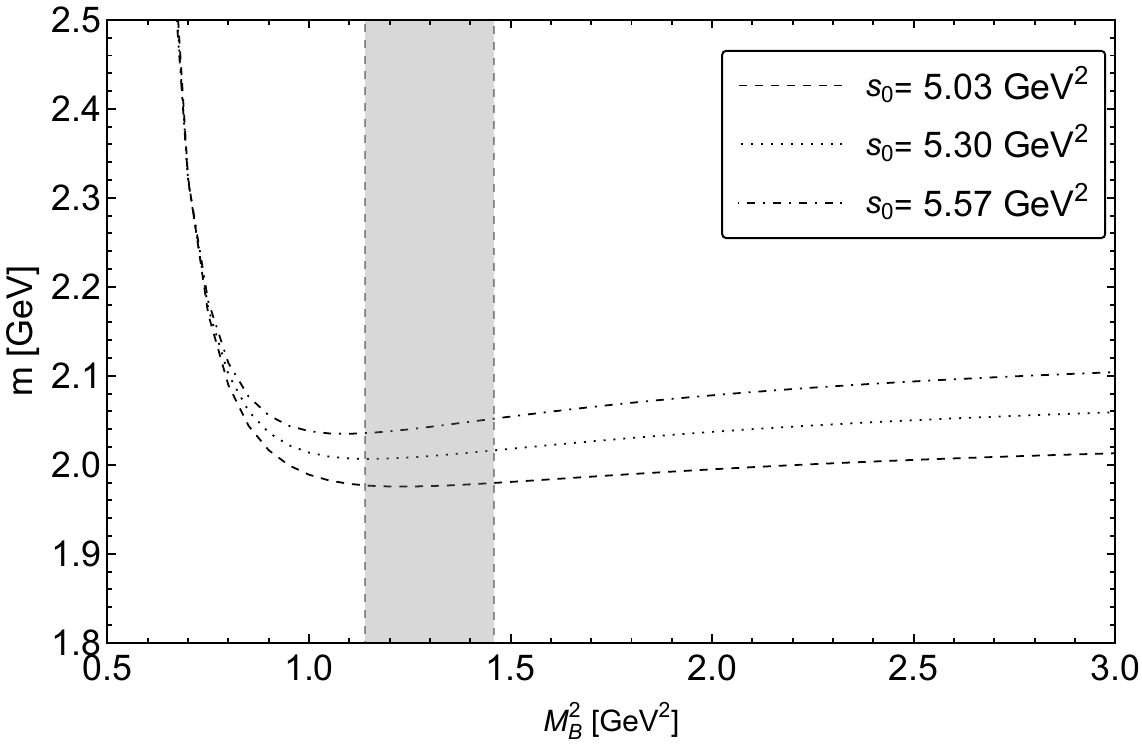}\label{fig:J7-MB-m-plus}}\\
  \caption{Variations of hadron mass with (a) continuum threshold $s_{0}$ and (b) Borel parameter $M_{B}^{2}$ for the hybrid baryon with $I(J^{P})=1/2(1/2)^{+}$ from $J_{7}$.}
  \label{fig:Result-1}
\end{figure}

As an example, we show our numerical analysis for the current $J_{7}$, which can couple to both negative- and positive-parity hybrid baryons with $I(J)=1/2(1/2)$. We firstly investigate the hybrid baryon with $J^P=1/2^{+}$. We require $R_{D>6}\le 10\%$ to guarantee the good OPE convergence, leading to the lower bound of the Borel mass $M_{B}^{2}\geq 1.14~\mathrm{GeV}^{2}$. By using this value and requiring that the pole contribution be larger than $30\%$, the threshold value $s_{0}$ can be determined around $5.3~\mathrm{GeV}^{2}$. We set $5\%$ uncertainty of $s_{0}$ in our analyses.
In Fig.~\ref{fig:J7-s0-m-plus}, we show the variations of the hybrid baryon mass $m_{X}$ with $s_{0}$ for various Borel mass $M_{B}^{2}$, from which the optimal working region is $5.03\leq s_0\leq 5.57~\mathrm{GeV}^{2}$. Using the central value of $s_{0}$, the upper bound of $M_{B}^{2}$ can be obtained by requiring the pole contribution be larger than 30\%. Finally, the Borel parameter can be determined to be $1.14\leq M_{B}^{2}\leq 1.46~\mathrm{GeV}^{2}$. 

As shown in Fig.~\ref{fig:J7-MB-m-plus}, the Borel curves are very stable in the working parameter regions so that the mass predictions are reliable. The mass of the hybrid baryon with $I(J^{P})=1/2(1/2)^{+}$ is predicted as
\begin{equation}
  \begin{aligned}
    &M_{+}=2.01_{-0.08} ^{+0.12}~\mathrm{GeV}\, ,
  \end{aligned}\label{J7-plus-mass}  
\end{equation}
where the errors come from $s_{0}$ and various condensates in Eq.~\eqref{parameters}. 

\begin{table}[t!]
  \caption{Mass spectra for all hybrid baryons. In the last column, the hadron masses are extracted by omitting the contributions of dimension-eight condensates. The notation ``-'' means no reliable mass prediction in this channel. The last column gives the results without considering the contributions of dimension-eight condensates.}
\renewcommand\arraystretch{1.2} 
  \setlength{\tabcolsep}{0.3em}{ 
    \begin{tabular}{cccccccc}
      \hline\hline  
      $J_i$ & $N/\Delta$ & $s_0[\mathrm{GeV}^2]$ & $M_B^2[\mathrm{GeV}^2]$ & $M_X$ [GeV] & $PC[\%]$ & $\tilde M_X$ [GeV] \\
      \hline 
        \multirow{2}{*}{$J_1$} & $N_{1/2^{-}}$ &  $12.4(\pm 5\%)$ & 2.75-2.82 & $3.15_{-0.06}^{+0.06}$  & 30.9  &$3.30_{-0.06}^{+0.06}$  \\
      &  $N_{1/2^{+}}$& - & - & - & - &  -\\   
      \cline { 2 - 7 }\multirow{2}{*}{$J_2$ }& $\Delta_{1/2^{-}}$& $16.5(\pm 5\%)$ & 2.59-3.60 &$3.64_{-0.08}^{+0.09}$  &41.6 & $3.63_{-0.08}^{+0.09}$\\
      &  $\Delta_{1/2^{+}}$ & - & - & - & -   & -\\ 
     \cline { 2 - 7 }       \multirow{2}{*}{$J_3$} & $N_{1/2^{-}}$ &  $7.6(\pm 5\%)$ &1.66-2.05 & $2.43_{-0.06}^{+0.06}$  &37.0  & $2.14_{-0.05}^{+0.06}$\\  
   &    $N_{1/2^{+}}$ & $14.7(\pm 5\%)$ & 2.29-2.93  & $3.50_{-0.11}^{+0.17}$ &   38.3 &$3.29_{-0.13}^{+0.19}$ \\    
      \cline { 2 - 7 }
       \multirow{2}{*}{$J_4$} & $\Delta_{1/2^{-}}$ & $9.5(\pm 5\%)$ & 2.43-2.54 & $2.82_{-0.12}^{+0.20}$ & $21.1$ &$3.00_{-0.13}^{+0.19}$\\
      & $\Delta_{1/2^{+}}$&- & - & - & -&  -\\     
      \cline { 2 - 7 }
       \multirow{2}{*}{$J_5$} & $N_{1/2^{-}}$ &  $10.5(\pm 5\%)$ & 2.78-3.05 & $2.92_{-0.05}^{+0.05}$ & 22.7  &$2.77_{-0.05}^{+0.05}$\\\
       & $N_{1/2^{+}}$& - & - & - & -   &  -\\    
      \cline { 2 - 7 }
       \multirow{2}{*}{$J_6$} & $\Delta_{1/2^{-}}$ & - & - & - & -   &-\\
      &  $\Delta_{1/2^{+}}$&  $22.0(\pm 5\%)$ & 3.35-3.70 &$4.18_{-0.09}^{+0.10}$ & 54.9 & $4.31_{-0.09}^{+0.10}$ \\      
      
    \cline { 2 - 7 }      
     \multirow{2}{*}{$J_7$} & $N_{1/2^{-}}$ &  $15.0(\pm 5\%)$ & 2.56-3.07 & $3.52_{-0.10}^{+0.13}$ & 36.1  &$3.48_{-0.11}^{+0.14}$\\  
   &     $N_{1/2^{+}}$   & $5.3(\pm 5\%)$ & 1.14-1.46 & $2.01_{-0.08}^{+0.12}$ & 38.7 &-\\    
   
    \cline { 2 - 7 }
    \multirow{2}{*}{$J_8$} & $\Delta_{1/2^{-}}$ &  $13.8(\pm 5\%)$ & 2.06-2.63 & $3.24_{-0.06}^{+0.07}$  &59.5  &$3.36_{-0.07}^{+0.07}$\\  
   &     $\Delta_{1/2^{+}}$ & - & - & - & - &  -\\

   \cline { 2 - 7 }\multirow{2}{*}{$J_{1 \mu}$} & $N_{3/2^{-}}$&  $17.6(\pm 5\%)$& 3.24-3.94 &$3.75_{-0.09}^{+0.11}$ & 36.9  & $3.75_{-0.09}^{+0.11}$\\
      &  $N_{3/2^{+}}$& - & - & - & - & -\\ 

  \cline { 2 - 7 }
  \multirow{2}{*}{$J_{2 \mu}$ }
  &$\Delta_{3/2^{-}}$  & - & - & - & - &\\
   & $\Delta_{3/2^{+}}$ &  $15.5(\pm 5\%)$ & 2.52-2.87 & $3.45_{-0.06}^{+0.06}$ & 55.5 & $3.58_{-0.07}^{+0.07}$\\

  \cline { 2 - 7 }\multirow{2}{*}{$J_{3 \mu}$} &$N_{3/2^{-}}$&  $13.1(\pm 5\%)$& 2.96-3.21 &$3.23_{-0.07}^{+0.08}$ & 32.8 &$3.11_{-0.07}^{+0.08}$ \\
      &  $N_{3/2^{+}}$& - & - & - & - & -\\
  
  \cline { 2 - 7 }\multirow{2}{*}{$J_{4 \mu}$ }& $\Delta_{3/2^{-}}$  & - & - & - & - &  $2.68_{-0.11}^{+0.14}$ \\
      &  $\Delta_{3/2^{+}}$ & - & - & - & - & $3.49_{-0.09}^{+0.10}$ \\
      \hline\hline  
  \end{tabular}
}
\label{tab:results}
\end{table}
For all interpolating currents in Eq.~(\ref{Eq:currents}), the numerical analyses show that the OPE convergences and pole contributions may vary significantly for different channels. For instance, the pole contributions for the positive-parity channels from currents $J_{6}$ and $J_{2\mu}$  and negative-parity channels from currents $J_{8}$ are significant so that we set $PC\ge 50\%$ in the analyses. However, for the negative-parity channels from $J_{4}$ and $J_{5}$ the pole contributions are relatively small so that we set them as $PC\ge 20\%$.
Besides, the OPE convergence behavior for some channels, such as the positive-parity channel from $J_{2\mu}$, is much better than other ones. 
These channels can actually provide more reliable mass predictions. 

\begin{figure}[t!!]
\centering
\includegraphics[width=0.45\textwidth]{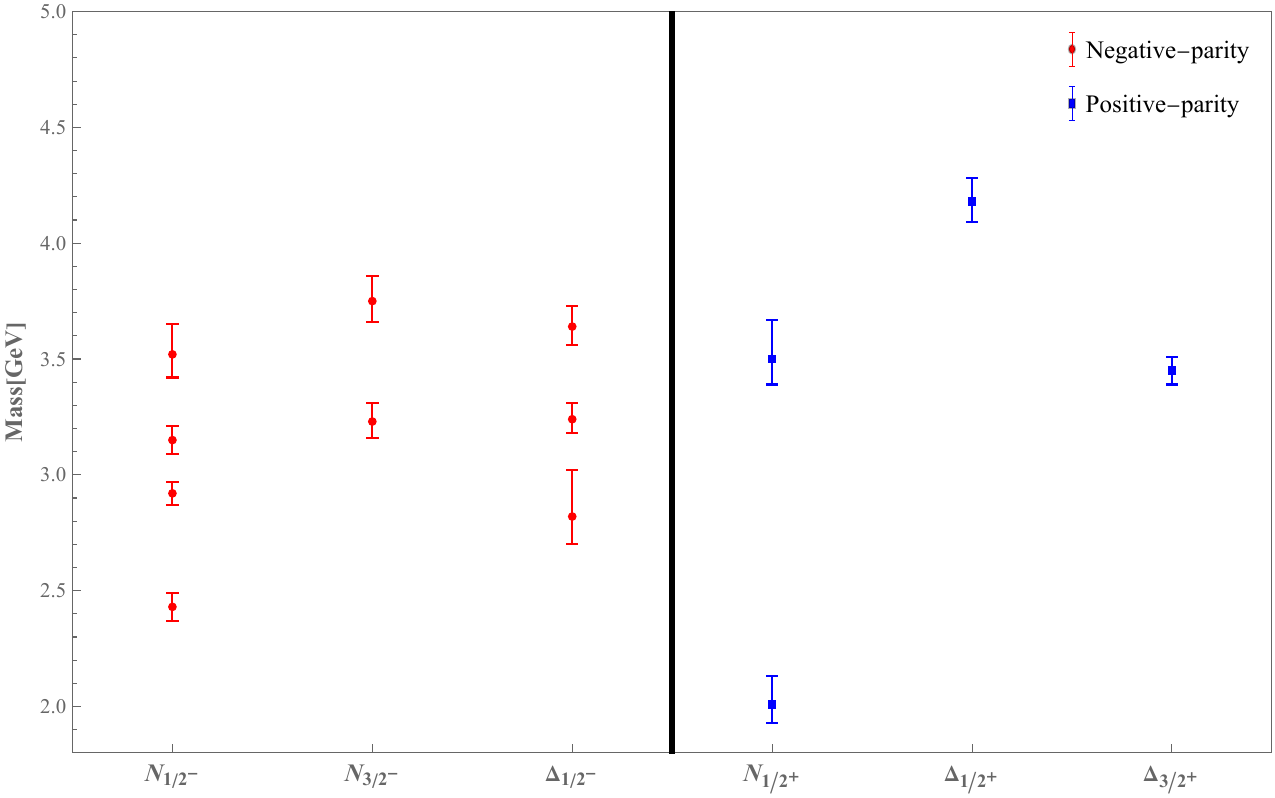}
\caption{Mass spectra for both negative-parity and positive-parity hybrid baryons.}
\label{fig:mass}
\end{figure}

In Table~\ref{tab:results}, we collect the numerical results for the hybrid baryons by using all interpolating currents in Eq.~\eqref{Eq:currents}. It is shown that there are reliable mass predictions for the positive-parity $N_{1/2^+}, \Delta_{1/2^+}, \Delta_{3/2^+}$ 
and negative-parity $N_{1/2^-}, N_{3/2^-}, \Delta_{1/2^-}$ hybrid baryons. 
For some currents, the QCD sum rules are unstable so that the mass predictions are unreliable for these channels, e.g., $N_{3/2^+}$ and $\Delta_{3/2^-}$, for which we denote as ``-'' in Table~\ref{tab:results}. As depicted in Fig.~\ref{fig:mass}, the lightest hybrid baryon is predicted to appear in the positive-parity $N_{1/2^+}$ channel, the lightest negative-parity $N_{1/2^-}$ is about $2.43~\mathrm{GeV}$ while the lightest $\Delta_{1/2^-}$ state is about 400 MeV higher. Such a mass splitting is almost the same with the mass difference $m_\Delta-m_N$ for the conventional baryons. For the other channels, the hybrid baryons are predicted to be above 3 GeV. 
Considering the renormalization scale at $\mu=2$ GeV, the predicted hybrid masses will be slightly increased without changing the overall spectrum structure. One may wonder that the traditional baryon states with the same quantum numbers also contribute to the correlation functions. Following the analysis of Ref.~\cite{Chen:2013zia}, we note that mixing with (lighter) conventional baryons that are weakly coupled to the hybrid currents can only raise the hybrid mass predictions of Table~\ref{tab:results}.

The accuracy of the factorization hypothesis for high-dimensional condensates in QCD sum rules remains unestablished. To quantify this uncertainty, we reanalyzed the mass sum rules omitting dimension-eight condensates and summarized the results in the last column of Table~\ref{tab:results}. For most channels, the $D=8$ contributions shift the predicted hadron mass by less than 10\%, consistent with findings in hybrid meson studies~\cite{Huang:2014hya}. Significant deviations occur, however, in the $N_{1/2^+}$ channel of $J_7$ and $\Delta_{3/2^\pm}$ channels of $J_{4\mu}$​, where the stability of the sum rules is qualitatively altered by the inclusion of these condensates.

\begin{figure}[t!!]
  \centering
\includegraphics[width=0.42\textwidth]{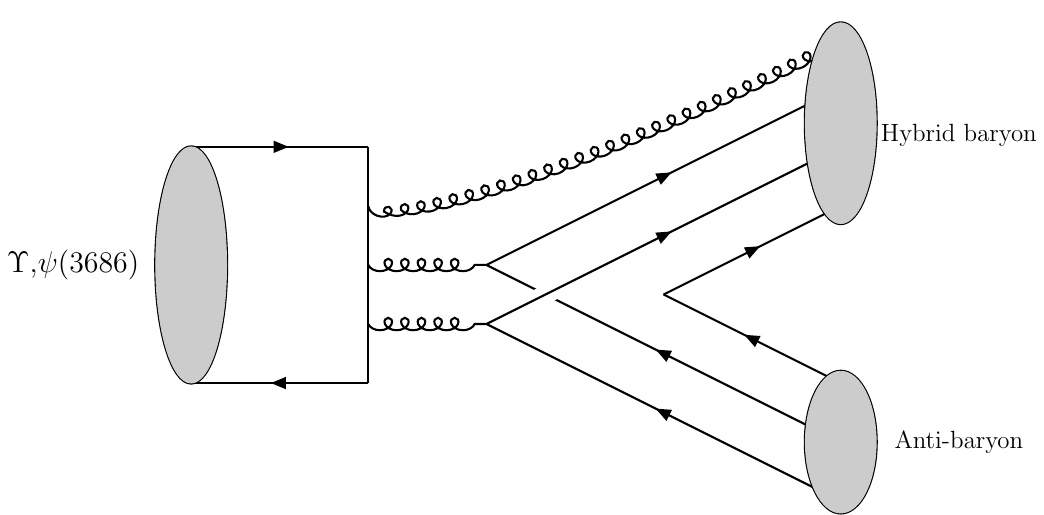}
  \caption{Possible production mechanisms of light hybrid baryons from $\Upsilon/\psi(3686)$ decays.}
  \label{fig: production}
\end{figure}

\begin{figure}[t!!]
  \centering
\includegraphics[width=0.45\textwidth]{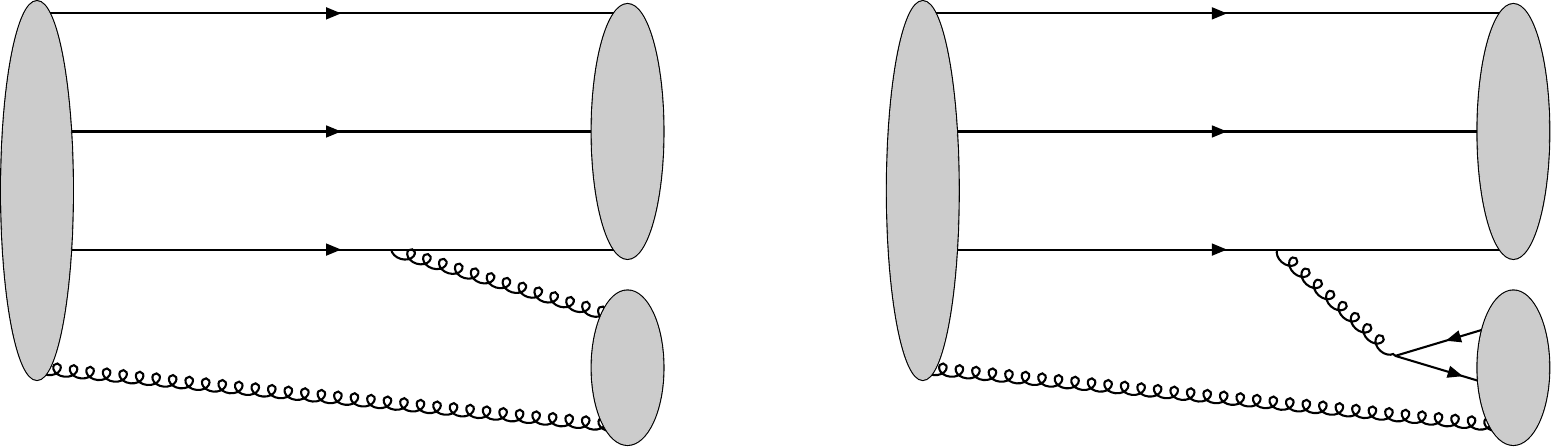}
  \caption{Possible decay mechanisms of light hybrid baryon to gluon-rich channels.}
\label{fig:decay}
\end{figure}

{\it Production and decay mechanisms.}---The decay of charmonium and bottomonium mesons can provide an ideal platform for producing gluon-rich hadron states. For instance, the well known $1^{-+}$ hybrid candidate $\eta_{1}(1855)$ was observed in the radiative decays of $J/\psi$~\cite{BESIII:2022riz,Chen:2022isv}. Similarly, the charmonium/bottomonium decay processes may produce light hybrid baryons, if kinematically allowed. According to our results, all predicted hybrid baryons can be generated through the  bottomonium decays via the three-gluon emission processes, such as $\Upsilon\to\bar{N}N_{hyb}/\bar{\Delta}\Delta_{hyb}+c.c.$. Besides, the lightest $N_{1/2^{+}}$ hybrid state can also be generated through the charmonium decays $\psi(3686)\to \bar{N} N_{1/2^{+}}+c.c.$. We propose such  hybrid baryon production mechanisms in Fig.~\ref{fig: production}.

Carrying the same quantum numbers, the hybrid baryons could always mix with traditional baryon states, so that their decay channels to the conventional baryon plus meson are similar to the latter ones, as shown in Table~\ref{tab:decay_mode}. 
Moreover, a hybrid baryon may decay into final states containing gluon-rich hadrons such as hybrid mesons or glueballs, as depicted in Fig.~\ref{fig:decay}. Such characteristic decays may be useful to distinguish hybrid baryons from conventional states and pentaquark states. 
\begin{center}
\begin{table}[h!]\caption{Some possible decay modes into conventional baryon-meson final states for light hybrid baryons.}
  \renewcommand\arraystretch{1.2} 
  \setlength{\tabcolsep}{0.1em}{ 
\begin{tabular}{cccc}
\hline\hline
    Current   ~  & Hybrid   &Decay Modes \\ \hline 
 \multirow{1}{*}{$J_{1}$} 
 & \thead{$N_{1/2^{-}}$}
 &\thead{$N\pi/\eta^{(\prime)}/\rho/\omega$,\,$N(1440)\pi/\eta^{\prime}/\rho/\omega$,\,\\$\Delta\rho$,\,$\Delta(1600)\rho$,\,$N(1535)f_{0}(980)/a_{0}(980)$}\\
\cline{2-3}
\multirow{1}{*}{$J_{2}$} 
   &\thead{$\Delta_{1/2^{-}}$}
   &\thead{$N\pi/\rho$,\,$N(1440)\pi/\rho$,\,\\$\Delta(1232)\rho$,\,$N(1535)a_{0}(980)$}\\
 \cline{2-3}
\multirow{1}{*}{$J_{3}$} 
 & \thead{$N_{1/2^{-}}$}
 &\thead{$N\pi/\eta^{(\prime)}/\rho/\omega$,\,$N(1440)\pi/\eta^{(\prime)}$,\,$\Delta(1232)\rho$}\\
 &\thead{$N_{1/2^{+}}$}
 &\thead{$Nf_{0}(980)/a_{0}(980)$,\,$\Delta(1232)\pi/\rho$,\,\\$N(1440)f_{0}(980)/a_{0}(980)$}\\\cline{2-3}
\multirow{1}{*}{$J_{4}$} 
& \thead{$\Delta_{1/2^{-}}$}
 &\thead{$N\pi/\rho$,\,$N(1440)\pi/\rho$,\,$\Delta(1232)\rho$,\,$N(1535)a_{0}(980)$}\\
 \cline{2-3}
\multirow{1}{*}{$J_{5}$} 
 & \thead{$N_{1/2^{-}}$}
&\thead{$N\pi/\eta^{(\prime)}/\rho/\omega$,\,$N(1440)\pi/\eta^{(\prime)}/\rho/\omega$,\,\\$\Delta(1232)\rho$,\,$\Delta(1600)\rho$,\,$N(1535)f_{0}(980)/a_{0}(980)$}\\
 \cline{2-3}
\multirow{1}{*}{$J_{6}$} 
 &\thead{$\Delta_{1/2^{+}}$}
 &\thead{$Na_{0}(980)$,\,$N(1440)a_{0}(980)$,\,$\Delta(1232)a_{0}(980)f_{0}(980)$}\\\cline{2-3}
\multirow{1}{*}{$J_{7}$} 
 & \thead{$N_{1/2^{-}}$}
&\thead{$N\pi/\eta^{(\prime)}/\rho/\omega$,\,$N(1440)\pi/\eta^{\prime}/\rho/\omega$,\,\\$\Delta(1232)\rho$,\,$\Delta(1600)\rho$,\,$N(1535)f_{0}(980)/a_{0}(980)$}\\
&\thead{$N_{1/2^{+}}$}
 &\thead{$Nf_{0}(980)/a_{0}(980)$,\,$\Delta(1232)\pi/\rho$}\\\cline{2-3}
 \cline{2-3}
\multirow{1}{*}{$J_{8}$} 
& \thead{$\Delta_{1/2^{-}}$}
 &\thead{$N\pi/\rho$,\,$N(1440)\pi/\rho$,\,$\Delta(1232)\rho$,\,$N(1535)a_{0}(980)$}\\
\cline{2-3}
\multirow{1}{*}{$J_{1\mu}$} 
 &\thead{$N_{3/2^{-}}$}
 &\thead{$N\rho/\omega$,\,$N(1400)\rho/\omega$,\,$\Delta(1232)\rho$}\\
\cline{2-3}
\multirow{1}{*}{$J_{2\mu}$} 
 &\thead{$\Delta_{3/2^{+}}$}
 &\thead{$Nb_{1}/\rho$,\,$N(1440)b_{1}/\rho$,\,$\Delta(1232)b_{1}/h_{1}/\rho$}\\\cline{2-3}
\multirow{1}{*}{$J_{3\mu}$} 
 & \thead{$N_{3/2^{-}}$}
 &\thead{$N\rho/\omega$,\,$N(1400)\rho/\omega$,\,$\Delta(1232)\rho$}\\
\cline{2-3}
\hline\hline  
\end{tabular}}
  \label{tab:decay_mode}
\end{table}
\end{center}

{\it Summary.}---We calculate the masses of the light hybrid baryons with quantum numbers $I(J^{P})=1/2(1/2^{\pm}),3/2(1/2^{\pm}),1/2(3/2^{\pm})$ and $3/2(3/2^{\pm})$ in parity-projected QCD sum rule method by constructing local interpolating currents. The correlation functions and spectral densities are evaluated up to dimension eight condensates. We establish stable mass sum rules for the positive-parity $N_{1/2^+}, \Delta_{3/2^+}, \Delta_{1/2^+}$ and negative-parity $N_{1/2^-}, N_{3/2^-}, \Delta_{1/2^-}$ hybrid baryons, so that the mass predictions in these channels are reliable. 
The lightest hybrid baryon is predicted as the positive-parity states $N_{1/2^+}$  around 2.01 GeV while the negative-parity states are heavier, which is consistent with the LQCD's prediction~\cite{Dudek:2012ag}. Such mass spectra may suggest the same transverse electric gluon excitation with $J_g^{P_g}=1^+$ in both hybrid baryon and meson systems~\cite{Dudek:2012ag}. 
All other hybrids in our results are predicted to be above 3 GeV. These hybrid baryons may mainly decay into the conventional baryon-meson final states. We suggest hunting  for such hybrids through the gluon-rich processes in BESIII and BelleII experiments. 

{\bf ACKNOWLEDGMENTS:}
This work is supported by the National Natural Science Foundation of China under Grant No. 12305147, No. 12175318, 12575153 and No. 12075019, the Natural Science Foundation of Guangdong Province of China under Grants No. 2022A1515011922. TGS is grateful for research funding from the Natural Sciences \& Engineering Research Council of Canada (NSERC). JH was supported by the Mitacs Globalink Research Award and is grateful for the support and hospitality of SYSU.



\end{document}